# Designing of super-elastic freestanding ferroelectric thin films guided by phase-field simulations

Changqing Guo[1,2], Houbing Huang[1,2]

[1]*School of Materials Science and Engineering, Beijing Institute of Technology, Beijing 100081, China.*
[2]*Advanced Research Institute of Multidisciplinary Science, Beijing 100081, China.*

**Correspondence to:** Prof. Houbing Huang, School of Materials Science and Engineering & Advanced Research Institute of Multidisciplinary Science, Beijing 100081, China. E-mail: hbhuang@bit.edu.cn

**Abstract**

Understanding the dynamic behavior of the domain structure is critical to the design and application of super-elastic freestanding ferroelectric thin films. The phase-field simulation is currently a powerful tool for observing, exploring, and revealing domain switching behavior and phase transition in ferroelectric materials at the mesoscopic scale. The present review summarizes the recent progress of phase-field methods in the theoretical interpretation, mechanical response, and domain structure evolution of freestanding ferroelectric thin films, wrinkled structures, and nano-springs. Furthermore, the strong coupling relationship between strain and ferroelectric polarization in super-elastic ferroelectric nanostructures is confirmed and discussed, which brings new design strategies for strain engineering of freestanding ferroelectric thin film systems. To further promote the innovative development and application of the freestanding ferroelectric thin film system, the review ends with a summary and an outlook on the theoretical model of the freestanding ferroelectric thin film.

**Keywords:** Freestanding ferroelectric thin films, super-elastic, mechanical structure, topological domain structure, phase-field simulations

## INTRODUCTION

The advancement of semiconductor materials and micro-/nanofabrication technologies has promoted the growth of the modern electronics industry. Various materials with excellent qualities have been designed and prepared gradually, which meet the needs of the automated and intelligent manufacturing industries. Particularly in recent decades, Silicon-based complementary metal-oxide-semiconductor (CMOS)-centric design procedures play a critical role in the modern electronics industry. However, the emergence of flexible electronic devices has presented new challenges for the traditional CMOS processing technique and new opportunities for the existing electronic manufacturing industry[1–3]. The flexible electronic materials can be applied to fabricate flexible displays[4–6], nano-sensors[7–9], health monitors[10–12], and other electronic devices for intelligent processing and human-computer interaction[13]. For example, flexible electronic materials can integrate cutting-edge technologies, including intelligent signal processing, real-time information sensing, and nano-generating, to improve the intelligence level of biocompatible electronics. They are playing an increasingly critical role in the intelligent monitoring of human health and biomedical applications, which will significantly change the future of healthcare and the relationship between patients and electronics. With the rise and combination of artificial intelligence and the Internet of Things (AI and IoT, i.e., AIoT), these electronic devices have an increasingly wide range of applications.

However, most traditional inorganic materials with excellent electronic properties show poor stability in complex stress environments, severely limiting their application prospects in flexible electronic devices. Therefore, finding and preparing basic materials with excellent electronic and flexible properties is one of the keys to developing the flexible electronics industry. Generally, two complementary approaches can be applied to obtain electronic materials with superior mechanical-electrical properties: 1) design and innovation of flexible materials by developing novel materials, including polymers[14,15], hydrogels[16,17], liquid metals[18,19], and freestanding films[20,21]; 2) structural optimization design by designing traditional high-performance electronic materials into appropriate mechanical structures[22,23], including wrinkled[24,25], origami[26,27], kirigami[28,29], and textile[30,31] structures.

Ferroelectric perovskite oxides are one of the indispensable basic materials in the modern electronics industry because of their abundant physical properties, extensive research value, and long-term practical application prospects. High-performance ferroelectric oxides are critical component materials in the current electronics industry because they have outstanding electrical properties, including ferroelectricity, piezoelectricity, pyroelectricity, and dielectricity. It is widely applied in high-efficiency memories[32–35], micro-sensors[36,37], high-frequency filters[38,39], energy harvesting systems[40,41], high-energy-density capacitors[42–45], ultrasonic medical treatment[46,47], and other related devices[48–51]. In addition, it is expected to be applied in the field of high-temperature superconductivity[52,53]. However, perovskite ferroelectrics are generally considered to be brittle and unbendable[54,55]. Therefore, if they can be transformed into flexible through the above two approaches, it will significantly accelerate the growth of the flexible electronics industry and occupy increasingly widespread applications, as shown in Figure 1. In addition, organic ferroelectric polymer materials[56–60] have excellent flexibility, such as poly(vinylidenefluoride) (PVDF) and its copolymers, which can achieve sizeable mechanical deformation such as stretching, bending, and twisting without being damaged. Therefore, in designing and processing flexible electronic materials, ferroelectric materials with flexibility and super-elasticity have received extensive attention due to their excellent mechanical and electrical properties.

In this review, we mainly focus on the super-elastic freestanding ferroelectric thin films. First, the current super-elastic freestanding ferroelectric films and their experimental characterization results are summarized. Subsequently, the origin of the super-elasticity of freestanding ferroelectric thin films is revealed and understood by phase-field simulations. Then, the designed mechanical structures based on super-elastic ferroelectric thin films, such as two-dimensional (2D) wrinkles and three-dimensional (3D) nano-springs, are introduced (Figure 1). Finally, outlooks and prospects for the super-elastic and flexible freestanding ferroelectric thin films are listed.

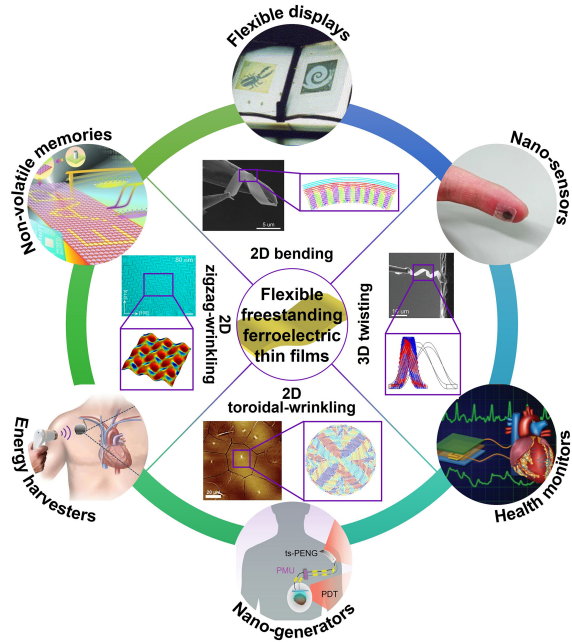

**Figure 1.** Novel mechanical structures[61–66] and applications[67–73] of super-elastic/flexible freestanding ferroelectric thin films.

## SUPER-ELASTIC FREESTANDING FERROELECTRIC FILMS

Due to the clamping effect of the substrate during the traditional process of preparing ferroelectric thin films, the ferroelectric thin films often show much worse electrical performance than those of freestanding materials, such as piezoelectric constant, switching speed, and switching voltage[74–77]. However, with the rapid development of new fabrication techniques for thin films[78], high-quality flexible freestanding ferroelectric thin films can be fabricated by a laser lift-off process[79–81], wet etching of sacrificial layers (e.g., $La_{0.7}Sr_{0.3}MnO_3$[82], $Sr_3Al_2O_6$[83–86], $BaO$[87], and $MgO$[88]), and mechanical exfoliation after deposition into 2D layered materials[89–93] (e.g., mica and graphene). As a result, the flexible ferroelectric thin films prepared by these methods have excellent mechanical elastic and electrical properties[94,95].

It is common to fabricate flexible freestanding thin films by etching sacrificial layers. The heterostructure of ferroelectric thin film and the sacrificial layer is first synthesized on $SrTiO_3$ (STO) substrate using pulsed-laser deposition or reactive molecular beam epitaxy. Then the sacrificial layer is etched using a specific solution. Finally, the freestanding thin film is transferred to a flexible substrate such as poly(ethylene terephthalate) (PET)[76,96,97] or polydimethylsiloxane (PDMS)[61,86,98–100]. Liu *et al.* fabricated freestanding single-crystal $BaTiO_3$ (BTO)[61] and $BiFeO_3$ (BFO)[99] ferroelectric thin films by pulsed-laser deposition (PLD) when using $Sr_3Al_2O_6$ as a sacrificial layer that was dissolved in water. As shown in the "2D bending" section of Figure 1, the fabricated BTO thin films can undergo a ~180° folding without any cracks in the in-situ scanning electron microscope (SEM) bending tests, which shows their super-elastic and flexible properties. Similarly, the BFO film can also achieve cyclic folding tests of up to 180°, and the largest bending strain during the bending process is observed to reach 5.42%.

Freestanding ferroelectric thin films provide an efficient and unique platform to study piezoelectric and flexoelectric effects in ferroelectric oxides. By PLD, Elangovan *et al.*[101] fabricated 30 nm thick flexible freestanding piezoelectric BTO thin films and showed they have good electromechanical-coupling properties. Under an external electric field, the thin film folds gradually and continuously by 180°, and the cycles of fold-unfold are reversible. Guo *et al.*[98] demonstrated the tunable photovoltaic effect in freestanding single-crystal BFO films and obtained multilevel photoconductance in BFO by altering the bending radius of the flexible device. This flexoelectric-control strategy confirms that the strain gradient can be used as a flexible degree of freedom to extend the functionality of flexible devices. In addition, Jin *et al.*[100] synthesized super-flexible freestanding $BiMnO_3$ membranes with stable ferroelectricity and ferromagnetism, which can maintain mechanical integrity under a nearly 180° folding.

## MODEL-GUIDED UNDERSTANDING OF SUPER-ELASTICITY IN FREESTANDING FERROELECTRIC THIN FILMS

The freestanding thin films demonstrate the advantages of tunable strain states compared with epitaxial growth on lattice-mismatched substrates. However, the source of the superior super-elasticity of freestanding films needs further exploration. The super-elasticity of freestanding ferroelectric membranes may originate from the mesoscale ferroelectric domain evolution in the presence of the external deformation; however, the direct observation of domain structure evolution of nanoscale freestanding films during continuous deformation is challenging by current experimental methods.

Figure 2 shows the behavior of domain structure evolution of freestanding ferroelectric thin films under bending deformation by phase-field simulations[61,102,103]. As shown in Figure 2B, during the continuous bending process of the freestanding BTO film, the mixed tensile and compressive stress generated by the bending (Figure 2A) causes the electric dipole to rotate continuously in the transition region connecting the *a* and *c* domains, forming "vortex-like" domain structures. The formation of "vortex-like" domains essentially eliminates the sharp stress caused by lattice mismatch, allowing the freestanding film to maintain mechanical integrity during the bending process. The continuous rotation of the electric dipole can be explained by phenomenological Landau theory; as seen from Figure 2C, four minima in the energy landscape exist in a bulk stress-free state of BTO, corresponding to two *a* domains and *c* domains, respectively. When in the bending state, the energy barrier between the *a* and *c* domains decreases, indicating that the polarization transition between the *a* and *c* domains becomes easier. Figure 2D shows that the "vortex-like" domains have an obvious size effect; the *a/c* phase with a vortex-like structure emerges only when the film thickness reaches 12 nm. As shown in Figure 2E, Peng *et al.*[103] investigated the dynamics domain evolution process of bent freestanding BFO films, and an exotic ferroelectric vortex is generated by local ferroelastic switching when the bending angle is higher than the critical value (e.g., 17.19° for thickness is 80 nm). Comparing the dynamic evolution of different energy densities during phase-field simulations (Figure 2F), the generation of the vortex is driven by the minimization of $f_{elastic}$. The generation of a local ferroelectric vortex and the ferroelectric polarization rotation can effectively promote the reduction of elastic energy and modulate the mechanical stress on the BFO film during bending, which contributes to the accommodation of the large deformation and the super-elasticity.

As mentioned above, the excellent mechanical elasticity of freestanding ferroelectric films can be attributed to the strong coupling effects between bending strain state and electric dipoles, including the continuous rotation of polarization, the emergence of polar vortex domains, and ferroelectric phase transitions.

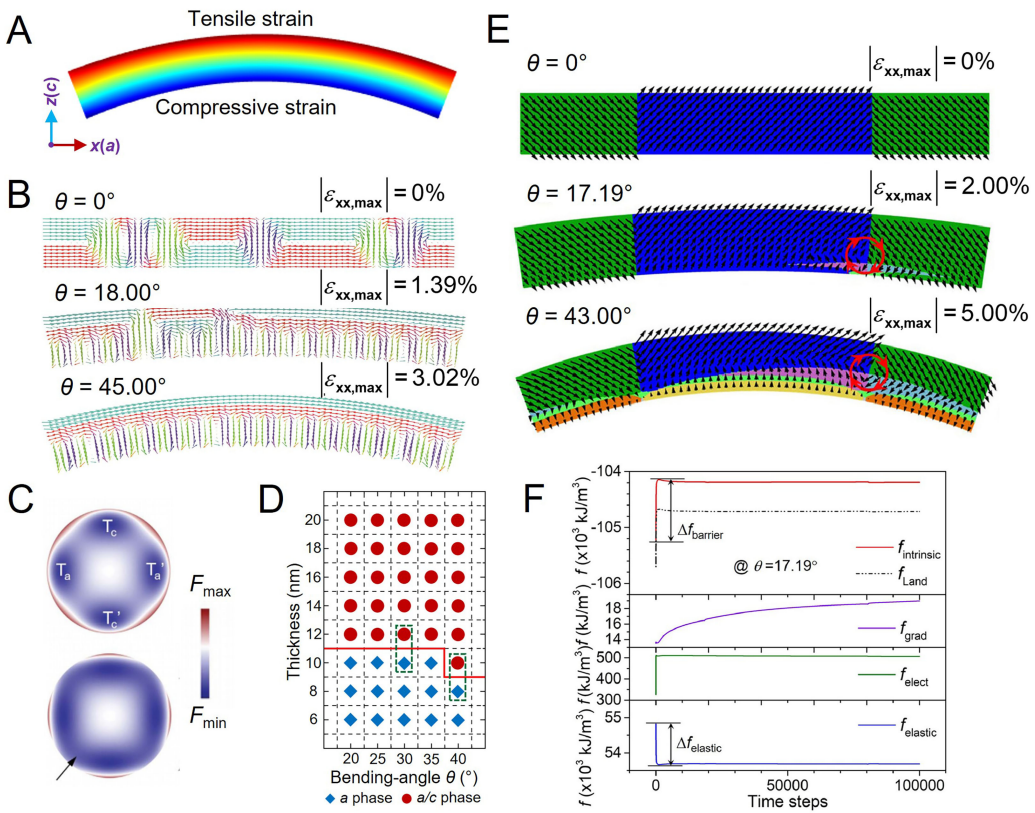

**Figure 2.** (A) Strain distribution of *n*-shape bent freestanding films[102]. (B) Domain structures and surface strain of the freestanding BTO thin film at different bending angles[102]. (C) Schematic illustration of the free energy landscape of bulk BTO (left) and freestanding thin films upon bending (right)[61]. (D) The size effect on domain patterns of the BTO thin film under bending[102]. (E) Dynamic evolution of ferroelectric domains in freestanding BFO thin films during bending[103]. (F) Dynamic evolution of the volume average energy density of freestanding BFO thin films under $\theta = 17.19°$ with time step[103].

Due to the absence of substrate constraints, freestanding ferroelectric thin films are an ideal medium for studying the distortion of crystal structures, physical properties, and prospective applications of ferroelectric materials. Therefore, the freestanding ferroelectric films are emerging as a new platform for exploring tunable strain states and novel mechanical structures. Chen et al.[104] systematically investigated the stability of 180° cylindrical domains in freestanding ferroelectric nanofilms under bending and explored the possibility of mechanical erasure of bit information ("0" and "1"). Ferroelectrics are potential candidates for data storage due to their switchable spontaneous polarization. Figure 3A shows a freestanding PbTiO$_3$ thin film divided into rectangular memory units, where the bit information "1" is represented by a cylindrical domain with downward polarization of radius $r$. From the phase diagram in Figure 3B, the bending deformation can effectively control the stability of the cylindrical domain and regulate the erasing of bit information. By changing the strain state of freestanding ferroelectric thin films, cylindrical domains can be erased, thereby providing a new theoretical idea for flexible memory devices. Besides bending structures based on super-elastic freestanding ferroelectric films, other mechanical structures, such as 2D wrinkled structures and 3D nano-spring structures, are also paving the way for novel flexible electronics.

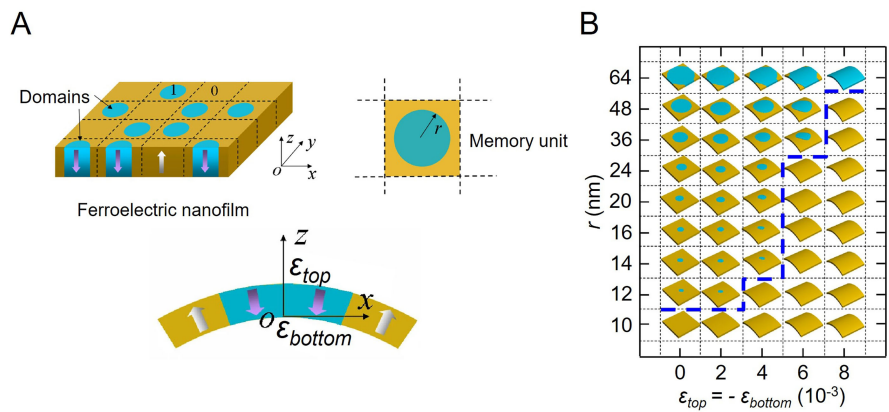

**Figure 3.** (A) Schematic illustration of a ferroelectric nanofilm with rectangular memory units (upper left), a single basic memory unit (upper right), and a bent single memory unit (bottom)[104]. (B) Phase diagram of the domain pattern in memory units under bending[104].

# FUNCTIONAL MECHANICAL STRUCTURES BASED ON SUPER-ELASTIC FERROELECTRIC THIN FILMS

## 2D wrinkled structure of freestanding ferroelectric thin films

The buckling instability mode, which leads to out-of-plane deformation, is prone to occur in thin-film structures under certain environmental stimuli (e.g., mechanical forces[105–109], temperature[110], van der Waals interactions[111], and localized diffusion of the solvent[112,113]). The occurrence of buckling modes may lead to structural and functional failure of the membranes, potentially limiting the performance of materials, and is often considered to be avoided. On the other hand, stress-driven buckling instability can self-assemble ordered surface topographies, such as sinusoidal, zigzag, labyrinthine, triangular, and checkerboard patterns. The physical properties of wrinkles show broad application prospects in the design of flexible electronic devices, the assembly of 3D complex microstructures[114], the morphology control of innovative optoelectronics and integrated systems[115–117], the measurement of mechanical properties of materials[118,119], and even medical assistance in diagnosis and treatment[120].

The formation of various wrinkled structures has been reported in thin film systems such as metals[105], graphene[111], organics[109,110,112,113], gels[121], and biological tissues[122], and more recently in freestanding ferroelectric thin film systems. As shown in Figure 4A, Dong et al.[63,64] successfully fabricated periodic wrinkle-patterned BTO/PDMS membranes based on the as-prepared super-elastic single-crystal freestanding BTO film. Moreover, the finely controlled wrinkle patterns, such as sinusoidal, zigzag, and labyrinthine patterns, have been obtained by changing the anisotropy and magnitude of the applied stress or adjusting the interfacial adhesion conditions. It is worth noting that the thickness of the super-elastic BTO film significantly affects the period of the wrinkle pattern. The thicker the BTO film, the larger the wavelength, which also has a corresponding theoretical explanation; that is, the wavelength and period of the wrinkle pattern are proportional to the thickness of the film[123–126]. Due to the peculiar morphology of the wrinkled ferroelectric thin film, its strain state is different from the interface mismatch strain of the traditional epitaxial thin film, resulting in a unique ferroelectric domain structure distribution. As shown in Figure 4B, modulation of the loading force induced by the scanning probe microscopy (SPM) tip can be switched to a periodic "braided" in-plane (IP) domain superstructure and opposite out-of-plane (OOP) domains between peaks and valleys. The unique domains depend on the strain state induced by the zigzag-wrinkled morphology and the loading of the SPM tip. Under the scanning force applied by an SPM tip, the zigzag wrinkle structure produces a "braided" IP strain state, which results in a unique distribution of domain structures. Therefore, the domain structure of freestanding ferroelectric wrinkled films can be modulated by the peculiar mixed strain state induced by the wrinkled morphology and the SPM tip. Figure 4C and D show the distribution of the corresponding strain and polarization components obtained from the phase-field simulation considering the flexoelectric effect.

As shown in Figure 4E, Cai et al.[127] observed a giant flexoelectric response at strain gradients up to ~ $4\times10^7$ m$^{-1}$ in a wrinkled structure based on high-quality flexible freestanding BFO and STO perovskite oxides[86]. Furthermore, the unusual bending-expansion and bending-shrinkage observed in bent freestanding BFO are also never seen before in crystalline materials. Corresponding theoretical models show that these novel phenomena are attributed to the combined action of flexoelectricity and piezoelectricity. These observations and theoretical models may provide a new path toward strain engineering and strain-gradient engineering of super-elastic freestanding ferroelectric thin films.

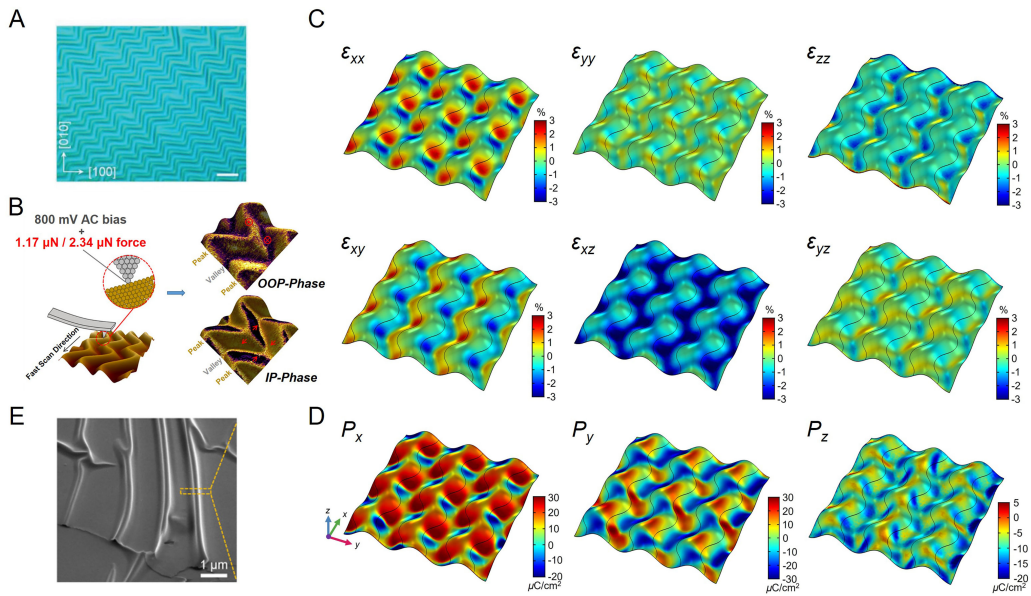

**Figure 4.** (A) Optical microscopy images of wrinkled BTO. The scale bar is 20 μm[63]. (B) Piezoresponse force microscopy (PFM) phase color images of zigzag-wrinkled BTO under scanning force induced by the SPM tip[64]. (C) Distribution of corresponding strain components of zigzag-wrinkled BTO by the SPM tip from phase-field simulations[64]. (D) Distribution of polarization components of zigzag-wrinkled BTO by the SPM tip[64]. (E) SEM image of wrinkled freestanding BFO[127].

In addition, there are many novel wrinkle structures in flexible ferroelectric polymer systems. As shown in Figure 5A, Guo et al.[65] found a periodic toroidal "target-like" wrinkle morphology in a flexible ferroelectric polymer poly(vinylidene fluoride-ran-trifluoroethylene) [P(VDF-TrFE)], as well as a peculiar toroidal topology topological texture using in-plane PFM (IP-PFM) measurements. Figure 5B-C show the in-plane strain and domain structure of wrinkled P(VDF-TrFE) films under an applied tensile strain of 7.3% from phase-field simulations. The distribution of the toroidal domain structure is related to the state of the mixture of tensile and compressive strains in the toroidal "target-like" wrinkled structure. According to this calculation, the electric toroidal moment[128] $G_z$ is much larger than $G_x$ and $G_y$, indicating the existence of in-plane toroidal order (Figure 5D). Moreover, the toroidal polar topology induces periodic absorption of polarized far-infrared waves, perhaps manipulating terahertz waves at the mesoscopic scale.

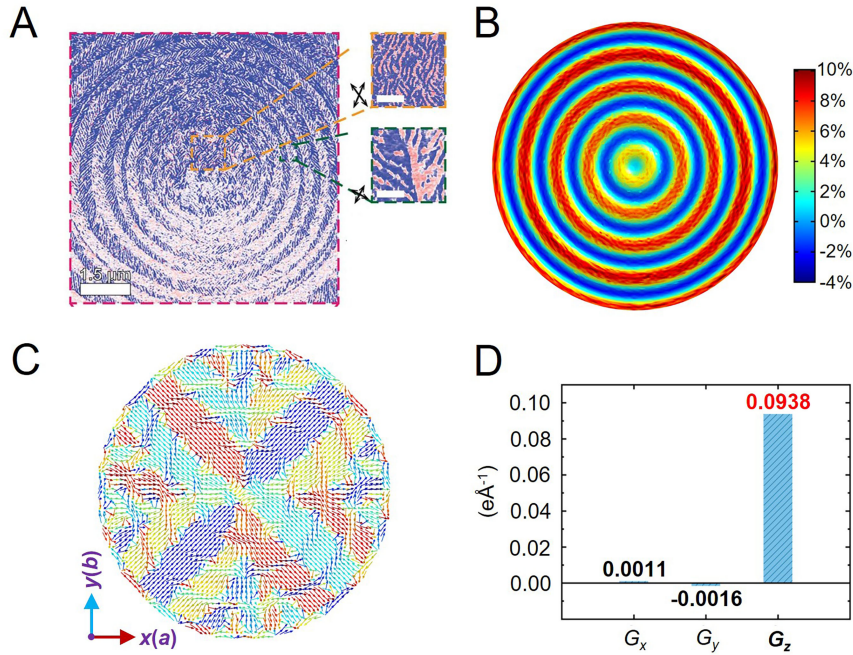

**Figure 5.** (A) IP-PFM phase image showing the toroidal polar topology of wrinkled P(VDF-TrFE) film. (B-C) The in-plane strain and domain structure of the wrinkled P(VDF-TrFE) under an applied tensile strain of 7.3% from phase-field simulations, respectively. (D) The electric toroidal moments correspond to the domain structure in Figure 5C.[65]

**3D nano-spring structure of freestanding ferroelectric thin films**

Ferroelectric materials display a range of individual polar topological states, including flux-closure domain[129,130], vortex[131–134], skyrmion bubble[135–137], meron[138], center domain[139], and sixfold vortex network[140], as a result of geometrical constraints between structural shape and material interface at the micro/nanoscale. The competition and coupling of elastic energy, electrostatic energy, and gradient energy can lead to the induction of novel polar topologies that are not stable in conventional bulk ferroelectrics under the combined influence of a sharply increased depolarization field caused by size effect, noticeable interface effect, and boundary constraint effect.

Under the demands of miniaturization and multi-functionalization of functional ferroelectric devices, more and more nanostructures with novel topological ferroelectric domains can be realized through top-down and bottom-up nanofabrication techniques. Based on the 2D Archimedes lattice, Shimada et al.[141] designed and studied 2D $PbTiO_3$ nanostructures including honeycomb-shape, kagome-shape, and star-shape through phase-field simulation. Furthermore, the polarization is gradually rotated at the junctions of different repeating units in these nano-metamaterials, forming a continuous flow pattern.

Among the unique nanostructures designed and produced, the emergence of nano-springs has received considerable attention. Nano-springs have the advantage of being able to achieve enormous amounts of mechanical deformation while maintaining mechanical integrity. This three-dimensional helical structure can have broad potential applications in electromechanical devices such as nanoactuators, nano-sensors, and nanomotors. Dong *et al.*[66] fabricated self-assembled $La_{0.7}Sr_{0.3}MnO_3$/$BaTiO_3$ (LSMO/BTO) ferroelectric nano-springs with excellent elasticity and recovering capability via a water-peeling off process. As illustrated in Figure 6A and G, the BTO nano-spring can be stretched or compressed to the geometric limit without breaking failure, achieving considerable scalability of 500%.

The phase-field simulations reveal that the excellent scalability originates from the continuously rotating ferroelastic domain structures, which provide displacement tolerance and energy to accommodate complex strains of mixed bending and twisting during mechanical deformation. Figure 6B and H show the strain component distribution $\varepsilon_{zz}$ of nano-springs under compression and elongation, respectively. This unique strain state results in the transition of 180° strip domains domains distributed around the surface of the ferroelectric nano-spring, as shown in Figure 6C and I. However, during the compression/ elongation process, the outer and inner surfaces of the nano-spring are subjected to opposite strains, which results in different ferroelectric domain structures on the two surfaces (Figure 6D-E and J-K).It is worth noting that the bending moment is the largest in the region farthest laterally from the centerline of the three-dimensional helix (i.e., the middle region of a single helical structure); therefore, the strain effect in this region is more evident. For example, when the nano-spring is stretched, the compressive strain on the outer surface of the middle region is more evident than the tensile strain on the inner surface, so the electric dipoles in this region deflected perpendicular to the *z*-axis, forming a 180° domain structure in the out-of-plane of the BTO layer.

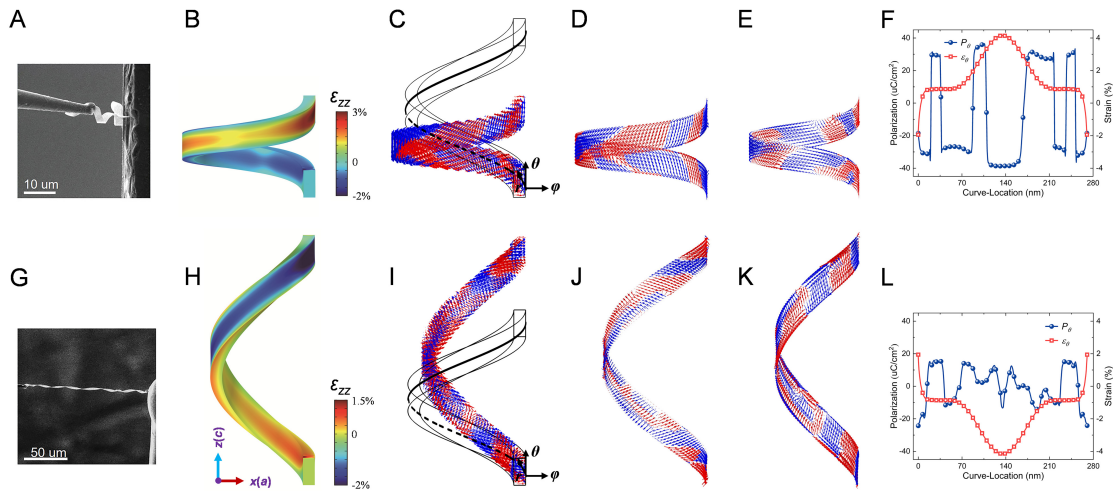

**Figure 6.** (A and G) In situ SEM images of the BTO nano-spring in compressive and tensile deformation, respectively. (B-F and H-L) Distribution of the strain component $\varepsilon_{zz}$, the domain structure, the outer surface domain, the inner surface domain, and the polarization component $P_\theta$ and strain distribution $\varepsilon_\theta$ along the middle line of the outer surface of the BTO nano-spring under compressive process and elongation process, respectively.[66]

Figure 6F and L show the relationship between the polarization and the strain distribution on the outer surface of the BTO nano-spring under compression and elongation, respectively. When the BTO thin films are twisted into the self-assembled nano-springs, the electromechanical coupling behavior differs from the direct relationship between polarization and strain in the previous ferroelectric film state (Figure 7). Therefore, the nano-springs design provides a novel conceptual framework and platform for strain engineering of freestanding ferroelectric thin films.

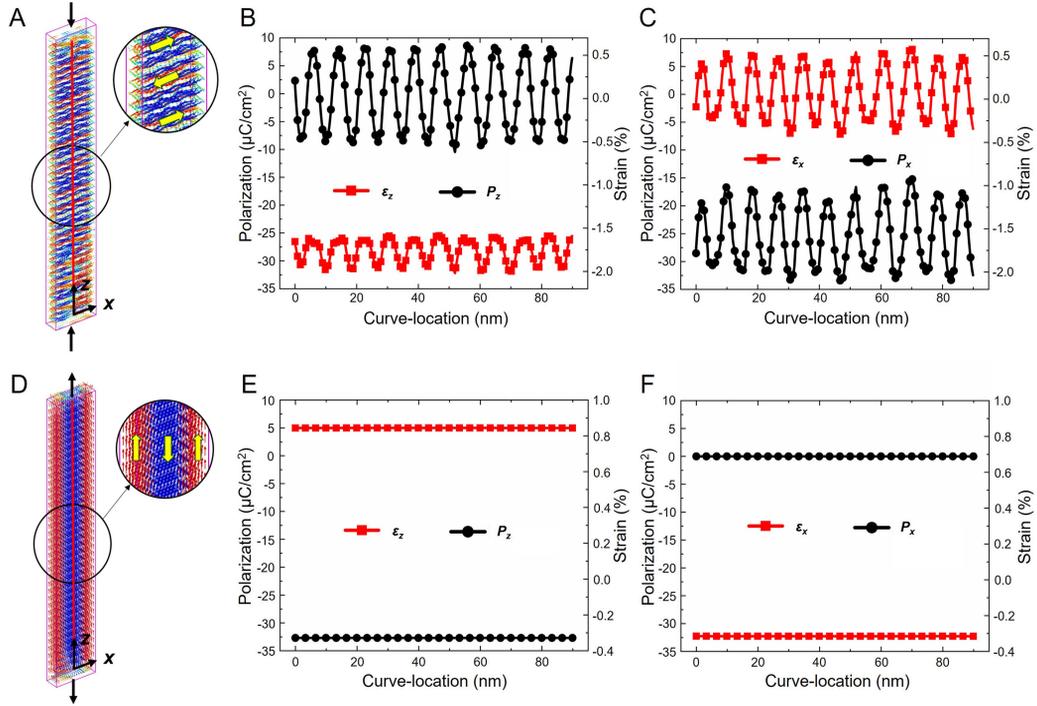

**Figure 7.** (A and D) The polarization distribution of BTO freestanding thin films under compressive and tensile deformation, respectively. (B-C and E-F) The polarization-strain distribution on the centerline of BTO freestanding thin films under compressive process and elongation process, respectively.[66]

## CONCLUSION AND OUTLOOK

As reviewed in this article, super-elastic ferroelectric materials are expected to be applied in a wide range of flexible electronic devices such as flexible memories, nano-sensors, and nanogenerators. It requires that ferroelectric materials maintain stable electrical performance under various mechanical deformations such as stretching, compressing, bending, and twisting, which involves the dynamic behavior of mesoscopic domain structures in ferroelectric materials. However, the direct observation of domain structure evolution of nanoscale freestanding ferroelectric films during continuous deformation is challenged by current experimental methods. Moreover, phase-field simulations have become increasingly critical in revealing the super-elasticity of freestanding ferroelectric films and the dynamical behavior of domain structures in different nanostructures. Therefore, the combination of experimental observations and phase-field simulations plays a key role in expanding the research and application of super-elastic ferroelectric materials in the field of flexible electronics.

To further play the guiding role of theoretical modeling in super-elastic freestanding ferroelectric thin films, the following crucial issues deserve attention and solutions:

(i) *The flexoelectric effect.* The flexoelectric effect describes the coupling between the electric polarization and strain gradient. Since the strain gradient is inversely proportional to the spatial scale (i.e., the gradient of the strain concerning the spatial coordinate), the flexoelectric effect is size-dependent[142]. Therefore, the flexoelectric effect becomes increasingly obvious and prominent as the size diminishes, and its contribution to the domain engineering of super-elastic freestanding ferroelectric thin films cannot be ignored. However, the theories and algorithms considering the flexoelectric effect in the phase-field model of super-elastic ferroelectrics still need further verification and improvement. For the freestanding ferroelectric system under mechanical deformation, it is necessary to establish a phase-field model considering the flexoelectric effect to study the origin, enhancement, and application of the flexoelectric effect on super-elasticity.

(ii) *Design, fabrication, and regulation of novel and functional mechanical structures.* In the regulation and application of freestanding ferroelectric systems, the design and preparation of novel and functional mechanical structures are also the focus of research. The influence of three-dimensional superstructures (e.g., isometric helicoids, kagome shapes, and hexagonal honeycombs) on the polarization distribution of ferroelectrics also requires systematic research and analysis.

(iii) *Responsive behavior in multiple fields.* Freestanding ferroelectric thin film systems exhibit several novel topological phenomena, such as skyrmion bubbles observed in freestanding multilayer thin films[136]. Therefore, it is essential to demonstrate the response and regulation of these topological structures in multi-physical fields (e.g., individual or mixed application of mechanical, electric, and magnetic fields). On the other hand, the response behavior of super-elastic freestanding ferroelectric nanostructures under the external field still needs further research and analysis. For example, the stretching and compressing behavior of ferroelectric nano-springs may be modulated by electric fields, which could have potential applications in the design of flexible nanorobots.

## DECLARATIONS

### Authors' contributions
Conceived and designed the manuscript: Huang HB, Guo CQ
Drafted and revised the manuscript: Huang HB, Guo CQ

### Availability of data and materials

Not applicable.

### Financial support and sponsorship

This work was financially supported by the National Natural Science Foundation of China (Grant No. 51972028) and the State Key Development Program for Basic Research of China (Grant No. 2019YFA0307900).

### Conflicts of interest

All authors declared that there are no conflicts of interest.

### Ethical approval and consent to participate

Not applicable.

### Consent for publication

Not applicable.